\documentclass[structabstract]{aa}  
\usepackage{graphicx}
\usepackage{natbib}
\usepackage{txfonts}
\begin{document}
   \title{Theoretical amplitudes and lifetimes of non-radial solar-like oscillations in red giants}

\author{M.-A. Dupret\inst{1,2} \and K. Belkacem\inst{2,1} \and R. Samadi\inst{1}
\and J. Montalban\inst{2} \and O. Moreira\inst{2} \and A. Miglio\inst{2} \and M. Godart\inst{2}
\and P. Ventura\inst{3} \and H.-G.~Ludwig\inst{4,5} \and A. Grigahc\`ene\inst{6} \and
M.-J Goupil\inst{1} \and A. Noels\inst{2} \and E.~Caffau\inst{5}}

\institute{
LESIA, Observatoire de Paris, CNRS UMR 8109, Universit\'e Paris Diderot, 5 place J. Janssen, 92195 Meudon, 
France \and
Institut d'Astrophysique et de G\'{e}ophysique de l'Universit\'{e} de
Li\`{e}ge, All\'ee du 6 Ao\^{u}t 17 - B 4000 Li\`{e}ge, Belgium \and INAF - Osservatorio Astronomico 
di Roma, MontePorzio Catone (RM), Italy \and 
CIFIST Marie Curie Excellence Team \and
GEPI, Observatoire de Paris, CNRS, Universit\'e Paris Diderot; 92195
Meudon Cedex, France \and 
Centro de Astrofisica da Universidade do Porto, Rua das Estrelas, 4150-762 Porto,
Portugal}
\abstract
{Solar-like oscillations have been observed in numerous red giants from
ground and from space. An important question arises: could we expect to 
detect non-radial modes probing the internal structure of these stars?}
{We investigate under what physical circumstances non-radial modes could be 
observable in red giants; what would be their amplitudes, lifetimes and heights in the
power spectrum (PS)?}
{Using a non-radial non-adiabatic pulsation code including
a non-local time-dependent treatment of convection, we compute
the theoretical lifetimes of radial and non-radial modes in 
several red giant models. Next, using a stochastic excitation
model, we compute the amplitudes of these modes and their heights 
in the PS.}
{Distinct cases appear.
Case~A corresponds to subgiants and stars at the bottom of the ascending giant branch.
Our results show that the lifetimes of the modes are mainly proportional to 
the inertia $I$, which is modulated by the mode trapping. The predicted amplitudes
are lower for non-radial modes. But the height of the peaks in the PS 
are of the same order for radial and non-radial modes as long as they can be resolved.
The resulting frequency spectrum is complex.
Case~B corresponds to intermediate models in the red giant branch.
In these models, the radiative damping becomes high enough to destroy the non-radial
modes trapped in the core.
Hence, only modes trapped in the envelope have significant heights in the PS
and could be observed. The resulting frequency spectrum of detectable modes
is regular for $\ell=0$ and $2$, but a little more complex for $\ell=1$ 
modes because of less efficient trapping.
Case~C corresponds to models of even higher luminosity.
In these models the radiative damping of non-radial modes is even larger than in the previous case 
and only radial and non-radial modes completely trapped in the envelope 
could be observed. The frequency pattern is very regular for these stars. The comparison between the 
predictions for radial and non-radial modes 
is very different if we consider the heights in the PS instead of the amplitudes.
This is important as the heights (not the amplitudes) are used as detection criterion.}{}

\keywords{Stars: oscillations -- Stars: interiors -- Convection}

\maketitle

\section{Introduction}

Oscillations of high radial order
p-modes have been  firmly detected in several red giants, 
from radial velocity data \citep{frandsen,bedding}
and photometric data from space  \citep{barban07,deriddernat}.
These oscillations are stochastically excited by turbulent motion in the outer layers 
of these stars. A particularly interesting case
is $\epsilon$~Oph, in which \cite{deridder06} detected equidistant frequencies 
from ground-based spectroscopic observations. \cite{hekker}  analysed the line-profile variations
and advocate the presence of non-radial modes. 
For the same star, \cite{barban07} used data obtained by
the MOST satellite. They detected at least 7 equidistant peaks and derived mode lifetimes 
of about 2.7 days.
\cite{kallingera} reexamined
the available data for $\epsilon$~Oph and claim to detect at least 21 independent modes
with longer lifetimes between 10 and 20 days. 
\cite{kallingerb} claim to have
detected non-radial modes with long lifetimes (more than 10 days)
in the star HD~20884 observed by MOST.
A very different result was found by \cite{stello06} who claim to have detected modes
with very short lifetimes in $\xi$~Hydrae. 
We finally mention the recent detection by the CoRoT satellite of 
solar-like radial and non-radial oscillations in many red giants \citep{deriddernat}. 
Different kinds of power spectra are found 
in these data of unprecedented quality, some of them showing regular patterns while others are more complex.
Modes with very long lifetimes (more than 50 days) are detected. 
  
From a theoretical point of view, an important study was made by \cite{dziem01} (hereafter D01).
They underline the mode trapping phenomenon present in these stars and the role
of radiative damping for several models with different luminosities. They predict that non-radial 
modes not trapped in the envelope would have much smaller
amplitudes than radial modes. 
\cite{houdek02} carried out theoretical computations of the amplitudes
and lifetimes of radial modes in the star $\xi$~Hydrae. They found lifetimes of 
the order of 15-20 days with a bump going up to infinite value (unstable mode)
around $110~\mu$Hz.
Finally, \cite{jcd} also discuss the mode trapping and inertia of 
non-radial adiabatic modes in red giants and subgiants and the possible consequences for the amplitudes.

In the present study, we compute theoretical lifetimes of radial and non-radial modes
of several red giant models, using a fully non-radial non-adiabatic pulsation code including
the non-local time-dependent treatment of convection by \cite{grig05} (G05). 
We discuss how these lifetimes are affected by
the mode inertia, the radiative damping in the core and the convective damping in the upper part of
the convective envelope. Using a stochastic excitation model \citep{Samadi00I,Samadi03a,Samadi03b}, we also compute the 
amplitudes of radial and non-radial modes and their theoretical heights
in the power spectrum (PS). As we show below, predictions for the heights are very different from
the amplitudes. Although non-radial modes in general have smaller amplitudes than
radial modes, this is not always the case for the heights. As the heights in the PS
are used as a criterion for the detection of frequencies, it is important to make this distinction.

In Sect.~\ref{energ}, we discuss the different energetic processes
determining the driving and damping of the modes: the radiative damping
in the core, the coherent interaction of convection with oscillations
and the stochastic excitation by turbulent motion
in the upper part of the convective envelope.
In Sect.~\ref{numeric}, we present the numerical tools, models and method used
to solve the problem.
In Sect.~\ref{results}, we present our results for specific red giant
models representative of the different cases that can occur.

\section{Energetic aspects of oscillations}
\label{energ}

Different mechanisms lead to the driving and damping of oscillations. 
For observed modes, two possibilities exist. Firstly, they could be 
self-excited. In such case the amplitudes are expected to grow until
a large amplitude limit cycle is reached. The observed amplitudes
of the high radial order p-modes in red giants are of several m/s. 
This is too small for such an interpretation 
and we reject this possibility here. Secondly, the modes can be damped, but 
turbulent motions supply them with energy in a stochastic way, allowing them
to reach observed amplitudes. 
We assume here that the high frequency modes of red giants 
are stochastic excited modes. The damping rate of the modes is given 
by an integral expression of this type:

\begin{equation}
\label{workeq}
\eta = -\frac{\int_V {\rm d}W}{2\,\sigma I\,|\xi_r(R)|^2 M}\,,
\end{equation}  

\noindent where we assumed the time-dependence $\exp(i\sigma t - \eta t)$ ($t$ is the time,
$\sigma$ is the angular frequency and $\eta$ is the damping rate).
$\int {\rm d}W$ is the work performed by the gas during one oscillation cycle.
$\vec{\xi}$ is the displacement vector, 
$I$ is the dimensionless mode inertia:
\begin{equation}
\label{ineq}
 I=\int_0^M |\vec{\xi}|^2\,{\rm d}m \:/\:  (M \,|\xi_r(R)|^2),
\end{equation}
$M$ is the total mass and
$\xi_r(R)$ corresponds to the radial displacement at the layer where the oscillations
are measured (Rosseland optical depth $\tau_R=0.1$ in our results).
In radiative zones, the work is easily modeled. It just results from the heat
given to the gas by radiation during each oscillation cycle. In convective zones,
the problem is much more complex and different terms can lead to damping: 
variations of convective flux, turbulent pressure, viscosity, and dissipation
of turbulent kinetic energy (G05). 
Stochastic excitation by turbulent motion also occurs,  leading to the observed amplitudes. This 
is discussed in Sect.~\ref{stoc}.

\subsection{Radiative damping}

In the g-mode cavity of red giants, the radial wave-number 
$k_r=\sqrt{\ell(\ell+1)}N/(\sigma r)$ becomes huge because of the
high density contrast between the core and the envelope
($N$ is the Brunt-V\"{a}is\"{a}l\"{a} frequency).
Large variations of the temperature gradient ensue, with loss of heat
in the hot phase, leading to radiative damping.  In the asymptotic limit, 
a simple expression for this damping can be obtained 
\citep{dziem77,vh98, godart}:

\begin{equation}
\label{numerator}
-\int_{r_0}^{r_c}\frac{{\rm d}W}{{\rm d}r}\,{\rm d}r\:\simeq\:\frac{K\,\left[\ell(\ell+1)\right]^{3/2}}{2\sigma^3}
\int_{r_0}^{r_c}
\frac{\nabla_{\rm ad}-\nabla}{\nabla}
\frac{\nabla_{\rm ad} N g L}{p\:r^5}\;{\rm d} r\:,
\end{equation}
where $r_0$ is the radius of the convective core (if present), $r_c$ is the upper radius 
of the g-mode cavity, $K$ is a normalization constant obtained by appropriate matching
with the envelope solution, $\nabla_{\rm ad}$ and $\nabla$ are the adiabatic and real gradients,
$g$ the gravity, $L$ the local luminosity and $p$ the pressure.
In the central regions of red giants, $N g/r^5$ is very high because of the high density contrast.
But at the same time, $p$ increases rapidly as we enter in the dense pure He core. As a compromise
between these two tendencies, the integrand of Eq.~(\ref{numerator}) reaches its largest values
around the bottom of the H-burning shell. The main radiative damping thus occurs in this  region
for red giants.
 
On the other side, the contribution of the g-mode cavity to the denominator of Eq.~(\ref{workeq})  
is simply:
\begin{equation}
\label{denominator}
8\pi\sigma\int_{r_0}^{r_c} |\vec{\xi}|^2\,r^2\rho\,{\rm d}r\;\simeq\;
4\pi\,K\,\int_{r_0}^{r_c} k_r\: {\rm d}r\:,
\end{equation}

\noindent which also can be significant because of the large $N$ 
($k_r\propto N$ in the core).
Eq.~(\ref{numerator}) scales like
$\tau_{\rm KH}^{-1}\,(R/R_c)^{7/2}\,(M/M_c)^{1/2}$, where 
$\tau_{\rm KH}=G M^2/(L R)$ is the Kelvin-Helmholtz time, $R_c$ is the radius
of the He core, $R$ is the total radius, $M_c$ is the mass of the He core
and $M$ is the total mass.
Equation (\ref{denominator}) scales like $(M_c/M)^{1/2}(R/R_c)^{3/2}$.
As the star climbs the red giant branch, $L$ and $R/R_c$ increase so that the 
numerator increases more quickly than the denominator and the radiative damping 
of non-radial modes increases.

\subsection{Coherent interaction with convection}
\label{TDCsec}

A very important region for the driving and damping of the modes is the transition
region where the thermal relaxation time is of the same order as the oscillation periods.
For solar-like oscillations, this transition region corresponds to the upper part
of the convective envelope. Moreover, in this region the time-scale of most  energetic turbulent
eddies is of the same order as the oscillation periods. Hence, the changing 
physical conditions due to the oscillations lead to periodic variations of the turbulent quantities
(convective flux, Reynolds stress, \ldots), which contribute to the work (positive or negative)
performed by the modes.
This is what we call the coherent interaction between convection 
and oscillations. It plays a major role in the damping of high radial order p-mode oscillations of red giants.

This interaction is difficult to model and only a few theories have been proposed
for linear oscillations. Two of them 
make use of a Mixing-Length 
formalism. 
The first by \cite{Gabriel1996} and further developed by G05 follows 
the original ideas of \cite{unno67}, where a turbulent viscous term 
opposite to the buoyancy is introduced.  
This treatment implemented in our non-radial non-adiabatic pulsation
code is used in this study. 
The second theory was developed by \cite{gough77} and follows the 
``kinetic of gas'' picture of the MLT. 
Both theories can include a non-local treatment \citep{spiegel,Gough77b,balmforth,dupnonloc06}. 
A  third formulation, no longer based on a mixing-length approach but on a Reynolds stress one 
was also proposed by \cite{xiong1997}. 
These theories encountered some successes, for example they obtain
the red edge of the instability strip \citep{houdek00,xiong01,dupinsta}. 
The reproduction of the Solar mode lifetime is also possible
but not easy \citep{balmforth,xiong00,dupretdamp06}. 

\subsection{Stochastic excitation}
\label{stoc}

For stochastically excited modes, 
the local squared amplitude of velocity variation at the layer where it is measured 
(we adopt the optical depth $\tau_R=0.1$ in our results) is given by 
\citep[e.g.,][]{Baudin05,Belkacem06b}:
\begin{equation}
\label{ampeq}
V^2\;=\;\frac{P}{2\,\eta\,M\,I}\;=\;\frac{\Pi}{2\,\eta\,M\,I^2} \, ,
\end{equation}
$P$ is the power stochastically supplied to the modes: 
\begin{equation}
\label{stocheq}
P\;=\;\frac{1}{8\,M\,I}\,(C_R^2+C_S^2)\,,
\end{equation}
where $C_R^2$ and $C_S^2$ are the turbulent Reynolds stress and entropy contributions,
respectively  \citep[see][for details]{Samadi00I,Belkacem06b}. Both play a significant role in red 
giants and are included in our study. 
$\eta$ and $I$ are the damping rate and inertia defined in Eqs.~(\ref{workeq}) and (\ref{ineq}), respectively.
To isolate the effect of inertia in Eq.~(\ref{ampeq}), we introduce the product $\Pi=P I$ which 
is, according to Eq.~(\ref{stocheq}), independent of inertia.

We also introduce the maximum height of the mode profile in the PS, which is an observable and will 
permit us to draw conclusions about the mode detectability (see Sect.~\ref{results} for details).
To this end, one has to distinguish between two cases, namely resolved and unresolved modes. 
The resolved modes present a Lorentzian profile in the PS and their heights are given by 
\citep[see e.g.][]{Chaplin05,Belkacem06b}:
\begin{equation}
\label{heighteqres}
H\;=\;\frac{V^2(R)}{\eta}\;=\;\frac{\Pi}{2\,\eta^2\,M\,I^2}\;=\;\frac{\Pi\tau^2}{2\,M\,I^2}\,
\end{equation}
where $\tau=1/\eta$ is the mode lifetime. 

When $\tau \gtrsim T_{\rm obs}/2$ ($T_{\rm obs}$ being the duration of observations), 
the modes are not resolved. In the limit  $\tau\to \infty$, the heights in the PS 
tend to behave like \citep[see e.g.][]{berthomieu2001,lochard05}:
\begin{equation}
\label{heighteqnores}
H_{\infty}\;=\;\frac{T_{\rm obs}}{2}\;V^2(R)\;=\;\frac{\Pi T_{\rm obs}\tau}{4\,M\,I^2}\,.
\end{equation}
In the theoretical predictions of the heights presented in this paper, we choose the very favorable case 
of the CoRoT long runs with 
$T_{\rm obs}=150$~days. We use Eq.~(\ref{heighteqres}) when $\tau\leq T_{\rm obs}/2=75$~days and
Eq.~(\ref{heighteqnores}) when $\tau> T_{\rm obs}/2=75$~days, so that
$H$ is a continuous function of $\tau$.

Finally, the observed velocity amplitudes are obtained by
integration of the projected local velocity over the visible stellar disk \citep{dziem77b}. 
This introduces a visibility factor depending on
the inclination angle of the star (through the factor $P_{\ell}^m(\cos i)$). 
We do not include this factor in our theoretical study (the inclination angle is unknown).

\section{Numerical tools, models and method}
\label{numeric}

\subsection{Structure models}
\label{struct_model}

The equilibrium models analyzed in this study were computed with  the
version updated for asteroseismology of the  code 
{\sc ATON} \citep{Mazzitelli79}. A detailed description of the numerical techniques and
implemented physics is given 
in \cite{ATON} and references therein. For the models 
in this paper we adopted  the classical mixing-length theory of convection
\citep{Bohm-Vitense}, 
with a value of the mixing-length parameter $\alpha_{\rm MLT}=1.8$, and we assume core 
overshooting that was treated as a diffusive mixing process. For the parameter
describing 
the extension of that extra mixing we took $\alpha_{\rm OV}=0.015$ 
\citep[see][for details]{ventura98},
and for the chemical composition $X=0.72$ and $Z=0.012$.
The radiative opacities  are those from OPAL \citep{Iglesias} for the 
metal mixture of \cite{GN93}
completed with \cite{alfer} at low temperatures, and the conductive
ones come  from \cite{Potekhin}.
Thermodynamic quantities are derived from  OPAL  \citep{rogers-nayf}, 
\cite{Saumon} for the pressure ionization regime and  \cite{Stolzmann96} treatment 
is used for He/C/O mixtures.
The nuclear cross-sections are from the NACRE compilation \citep{Angulo}, and
the surface boundary conditions
are provided by a grey atmosphere following the \cite{Henyey} treatment.
Turbulent pressure is not included in our structure models.

\begin{figure}
\includegraphics[width=8.5cm]{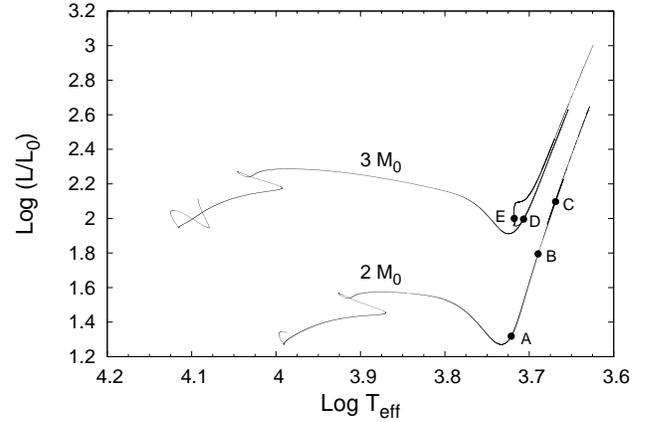}
\caption{HR diagram with the location of our five models.}\label{HR}  
\end{figure}

We focus our study on five well chosen models representative of the
different situations that can occur. Some of their global characteristics are given
in Table~\ref{models}. $\nu_{\rm cut}=(g/2)\,\sqrt{\Gamma_1\rho/p}$ is the cut-off 
frequency of an isothermal plane-parallel atmosphere
(estimated here at the photosphere). Evolutionary tracks with the location of these models
in the HR diagram are given in Fig.~\ref{HR}.

\begin{table}
\centering
\begin{tabular}{ccccccc} \hline
\noalign{\smallskip}
 & $M/M_{\sun}$ & $T_{\rm eff}$ & $\log(L/L_{\sun})$ & $\log g$ & $R/R_{\sun}$ & $\nu_{\rm cut}$  \\
 &  &  (K) & & & & ($\mu_{\rm Hz}$)\\
\noalign{\smallskip}
\hline
\noalign{\smallskip}
A &   2 &    5264  &  1.32 &  3.26  &  5.50  &  315 \\
B &   2 &    4892  &  1.80 &  2.65  &  11.0  &  82.1 \\
C &   2 &    4665  &  2.10 &  2.27  &  17.2  &  34.6 \\
\noalign{\smallskip}
D &   3 &    5091  &  2.00 &  2.70  & 12.8  & 88.6 \\
E &   3 &    5222  &  2.00 &  2.74  & 12.2  & 96.3 \\
\noalign{\smallskip}
\hline
\end{tabular}
 \caption{Global parameters of our models.}
  \label{models}
\end{table}

\subsection{Non-radial non-adiabatic oscillations}
\label{nonadsec}

We use the non-adiabatic pulsation code MAD \citep{duphd} to compute the theoretical 
lifetimes and inertia of the modes. 
We use the TDC treatment of G05 for the modelling of the convective flux and turbulent pressure
variations entering in the non-adiabatic equations. 
Different ingredients must be specified. We keep a local treatment for the convective flux.
The variations of the turbulent pressure and entropy gradient are treated in a non-local way, 
with values of the non-local parameters $b=3$ and $c=3.5$ according to the definition of 
\cite{balmforth}.
Our value of $b$ is derived from a fit of the
turbulent pressure in the subadiabatic atmospheric layers of a solar hydrodynamic simulation 
\citep{dupnonloc06}. 
We recall that our structure models are built without turbulent pressure.
We compute it a posteriori in order to be able to determine the turbulent pressure perturbation.
This implies that the effect of turbulent pressure on the stratification is not
taken into account in our models. However, the impact of this inconsistency 
on the theoretical lifetimes is expected to be 
negligible compared to others aspects.
The main source of uncertainties in any TDC treatment comes from the perturbation of the 
closure equations. In our TDC treatment, this uncertainty appears in the form of a 
free parameter $\beta$, as introduced by G05. We take by default the value $\beta=-2\,i$. 
Adopting this value for the Sun leads to predictions of the 
mode lifetimes in reasonable agreement with observations. It also gives in all our red giant
models stable high radial order p-modes, in agreement with our working hypothesis of stochastic
excitation. However, we discuss in Sect.~\ref{betachange} the impact of changing this value.

Some changes were required in the MAD code to address specific numerical difficulties
associated with red giants.
First, we changed the formalism to a Eulerian one in the core.
In the g-mode cavity of red giants where $\sigma << L_\ell$, we have $|p'/p| << |\delta p/p|$
and thus $\delta p/p\:\simeq\:(d \ln p/dr)\: \xi_r$. Hence, $\delta p$ and $\xi_r$ are not
independent functions and they do not allow us to capture the oscillatory behaviour of the eigenfunctions
in the cavity. Therefore, a Lagrangian formalism lacks numerical precision in the dense region of evolved
stars, which justifies our switch to a Eulerian formalism.

Another significant numerical difficulty comes from the very large number of nodes of 
the eigenfunctions in the g-mode cavity. \cite{vh98} and D01 used
a mixed treatment to address this difficulty: 
numerical solutions of the fully non-adiabatic equations in the envelope
are matched with asymptotic solutions in the g-mode cavity.
Our way to proceed is slightly different. Firstly we solve the fully non-adiabatic equations from the
center to the surface, which gives a first estimation of the lifetimes. 
Second, we separate the integrals in Eqs.~(\ref{workeq}) and (\ref{ineq}) in two parts: 
region 1 where $\sigma << L_\ell, N$ and region 2 (the remaining). 
In region 1 we use the results obtained in the asymptotic quasi-adiabatic limit
(Eqs.~(\ref{numerator}) and (\ref{denominator})), and in region 2 we use the 
fully non-adiabatic eigenfunctions obtained numerically in the first step. We then compare the 
results obtained by the two approaches: in many cases they are very close, showing 
that our fully numerical solution of the problem is precise enough. 
In other cases corresponding to high luminosity red giants, the number of nodes
becomes huge ($\gtrsim 1000$), which implies loss of numerical precision. In such cases we are more
confident in using results with the asymptotic treatment in the g-mode cavity and the fully non-adiabatic
solution in the rest of the star.

A last important numerical issue is associated with the mode trapping phenomenon, which is
discussed in detail in Sect.~\ref{trapsec}. Algorithms solving the non-adiabatic problem
converge towards the closest eigenvalue in the complex plane (the real part corresponds to 
the angular frequency $\sigma$ and the imaginary part to the damping rate $\eta$).
Some non-radial modes trapped in the envelope
have much lower inertia than the others. Hence, their damping rates are much higher
than for other non-radial modes, higher than the frequency separation between
consecutive non-radial modes: $\eta > \sigma_{n-1}-\sigma_n$. Hence, with a
real initial guess of the eigenvalues (e.g. the adiabatic frequencies),
it is impossible to converge towards the modes strongly trapped in the envelope. 
The only way is to adopt a complex initial value;
for example, interpolating the complex non-adiabatic eigenvalues of radial modes gives adequate
initial values, allowing the solution to converge towards the non-radial modes strongly trapped 
in the envelope.  

\subsection{Ingredients for the stochastic excitation models}

The power stochastically supplied to the modes, $P$ in Eqs.~(\ref{ampeq}) and (\ref{stocheq}), is 
computed as described in \cite{Belkacem08a}.  
The typical convective length-scales as well as the kinetic energy spectrum are poorly 
known for red giant stars. Given that the predictions of the stochastic excitation models strongly
depend on them, we infer both the kinetic energy spectrum and the injection length-scale from a
representative 3D numerical simulation. 
To this end, we computed a 3D radiation-hydrodynamical model atmosphere with the code
CO$^5$BOLD \citep{Freytag02,Wedemeyer04}. 
The simulation has a gravity $\log g=$ 2.5 and an effective temperature of $T_{\rm eff}=4960$~K. 
The model has a spatial mesh with $160 \times 160 \times 200$ grid
points, and a physical extent of the computational box of $573 \times 573 \times 243$~Mm$^3$.  
Using this 3D numerical simulation, we determine the injection length-scale 
in the layer where the driving is the largest 
\citep[see][for details]{Samadi03a}.
This estimation of the injection scale gives values larger than the 
mixing-length by a factor of 5-10 depending on the model. Using the mixing-length 
as the injection scale would give theoretical amplitudes much lower than typical observations. 
We assume here that the injection scale is constant in the driving region. In the 
case of the Sun, such approximation underestimates  by about $25\,\%$ the 
theoretical amplitudes \citep{Samadi03a}.
The models considered in our study have different
gravities. This affects the injection length-scale and we assume here that it scales  
as the pressure scale-height at the photosphere. Such a procedure has been verified in the case 
of $\alpha$ Cen A \citep{Samadi08a}. This allows us to compute the 
excitation rates for all the models presented in Sect.~\ref{struct_model}.
We use a Lorentzian profile for the  eddy time-correlation function \citep{Samadi03b}.
Eventually, concerning the $k$-dependency of the kinetic energy spectrum 
($k$ is the wavenumber in the Fourrier space of turbulence), we use 
the Broad Kolmogorov Spectrum (BKS) \citep{Musielak94} that reproduces the spectrum inferred from 
the 3D~simulation. 

\section{Results}
\label{results}

\begin{figure}
\includegraphics[width=8.5cm]{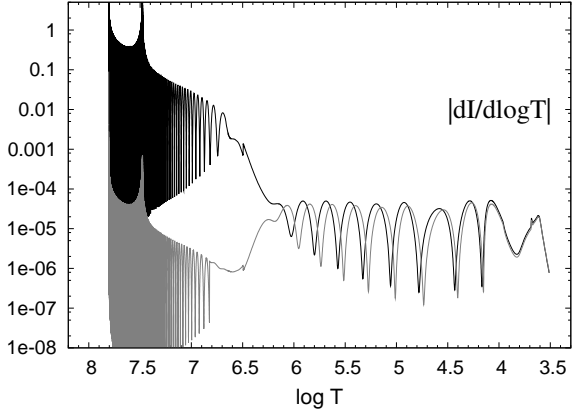}
\caption{Energy density $|{\rm d}I/{\rm d}\log T|$ 
for the mode $\ell=2$, $\nu=55.72\,\mu$Hz trapped in the envelope (grey) 
and the mode $\ell=2$, $\nu=53.87\,\mu$Hz trapped in the core (black), 
for model~B.}\label{in-B2}  
\end{figure} 

\begin{figure}
\includegraphics[width=8.5cm]{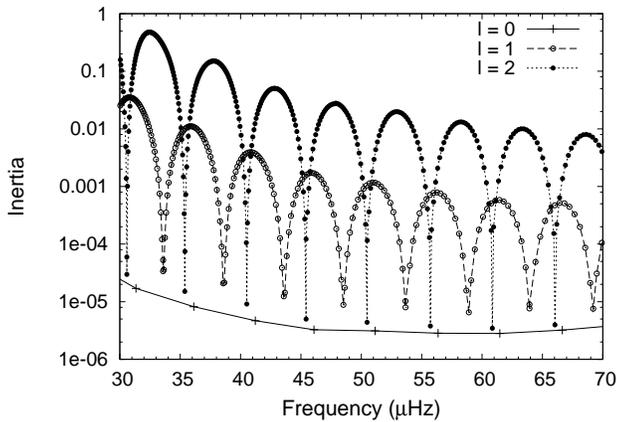}
\caption{Inertia of $\ell=0$, $1$ and $2$  modes of model~B.}\label{in-B1}  
\end{figure} 

\subsection{Mode trapping in evolved stars}
\label{trapsec}

Mode trapping is a very important phenomenon allowing to understand the behaviour
of non-radial modes in evolved stars. D01 and \cite{jcd} discussed it in 
the case of red giants. Here we recall basic aspects
that are important for the interpretation of our results. 
The Brunt-V\"{a}is\"{a}l\"{a} ($N$) and Lamb ($L_\ell$) frequencies are huge in the 
dense core of red giants. Because of these huge values, all non-radial 
modes with frequencies lower than the cut-off frequency and larger than the fundamental
radial mode have a mixed character. 
In the envelope, $\sigma^2 > L_\ell^2, N^2$ 
and they behave like acoustic modes (as in solar-like main sequence stars). 
But $\sigma^2 < L_\ell^2, N^2$ in the core 
and the same modes behave there like gravity modes. 
The presence of an evanescent region between the p- and g-cavities
is at the origin of the mode trapping. Some modes have significant
kinetic energy in the g-cavity and low ones in the p-cavity; they are
trapped in the core. Others have low kinetic energy in the g-cavity but high
ones in the envelope; they are trapped in the envelope.

An illustration of the 
distribution of energy of a mode trapped in the core and a mode trapped
in the envelope are shown in Fig.~\ref{in-B2}, where we give 
$|{\rm d}I/{\rm d}\log T|= -|\vec{\xi}|^2\,{\rm d}m/(M\,{\rm d}\log T)$.
Mode trapping directly affects the mode inertia. We give in Fig.~\ref{in-B1} 
the inertia as a function of frequency for $\ell=0-2$ modes. 
We first notice how dense the spectrum of 
frequencies for non-radial modes is, which is a direct consequence of the huge
Brunt-V\"{a}is\"{a}l\"{a} frequency in the core (according to the asymptotic theory, 
$P_{n, \ell}-P_{n-1,\ell}\simeq 2\pi^2/(\ell(\ell+1)\int N/r  \, {\rm d}r$).
Modes trapped in the envelope are the local minima with low inertia in Fig.~\ref{in-B1}.
The $\ell=2$ modes trapped in the envelope have
lower inertia than the $\ell=1$ modes trapped in the envelope, while it is the contrary
for all other modes. In agreement with D01 and \cite{jcd}, the size of the evanescent 
region and thus the efficiency of trapping increases with the degree $\ell$ of the modes.
This is very different from RR~Lyrae stars \citep{vh98} because
there the evanescent region is larger for $\ell=1$ in the observed range of frequency. 
It is very  interesting to notice that when only the modes trapped in the envelope
are considered, the frequency spectrum is similar to solar-type main 
sequence stars, with large and small separations appearing clearly.
The inertia appears directly in Eqs.~(\ref{workeq}), (\ref{ampeq}) and (\ref{heighteqnores}). 
Hence, the mode trapping is expected to strongly affect the amplitudes, lifetimes, and heights 
of the modes in the PS of red giants. 
In the next section, we present our main results for different models 
(Table~\ref{models}, Fig.~\ref{HR}).

\subsection{Model A: bottom of the red giant branch}

\begin{figure}
\includegraphics[width=8.5cm]{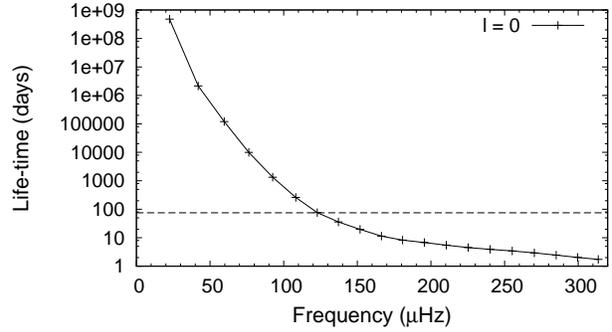}
\caption{Theoretical lifetimes of radial modes of model~A.
The horizontal line corresponds to our choice
of the resolution limit (75 days).}\label{life-A0}  
\end{figure} 

\begin{figure}
\includegraphics[width=8.5cm]{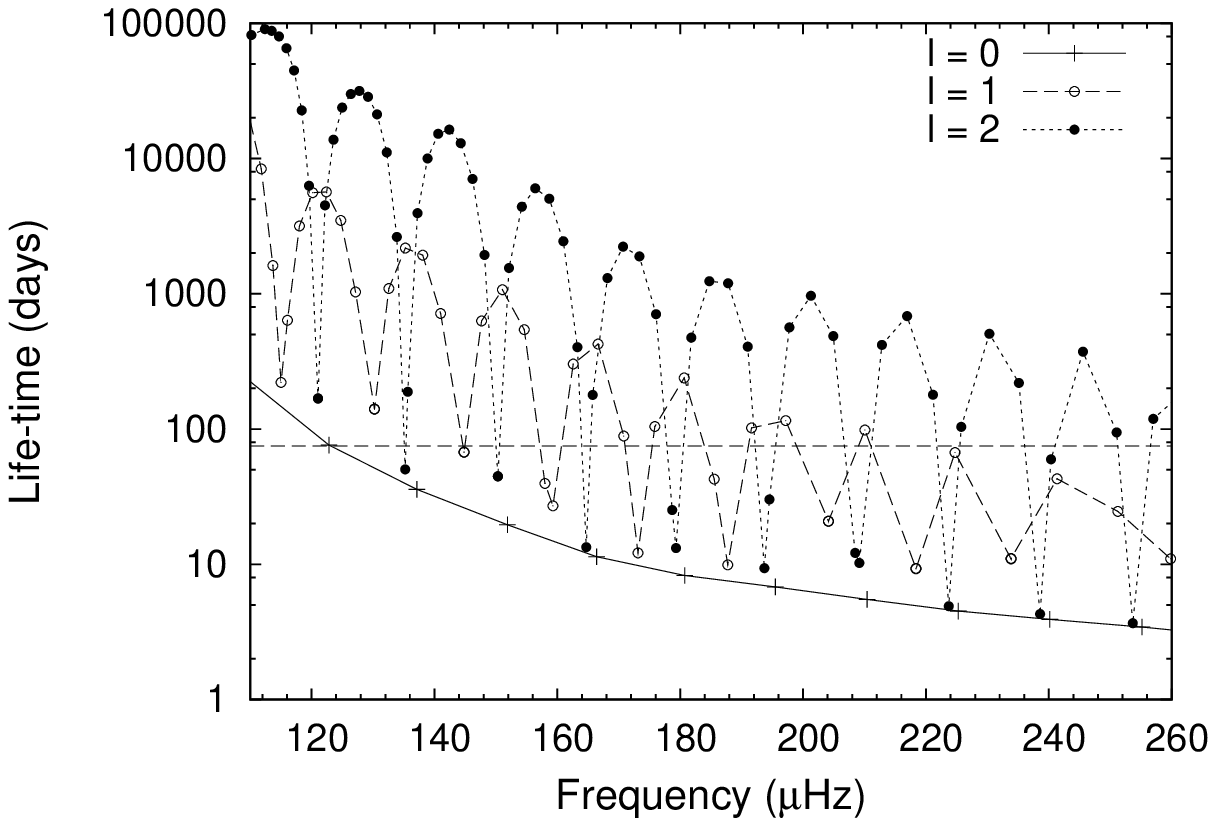}
\includegraphics[width=8.5cm]{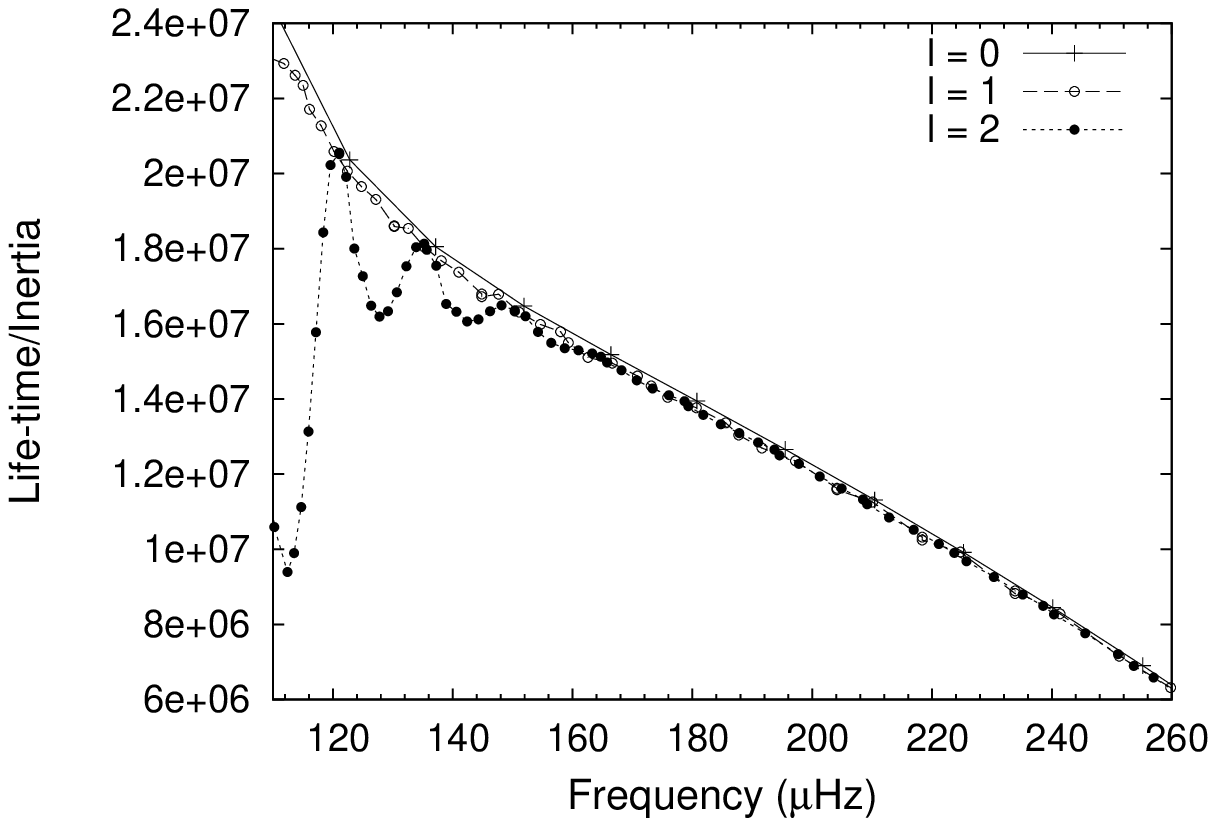}
\includegraphics[width=8.5cm]{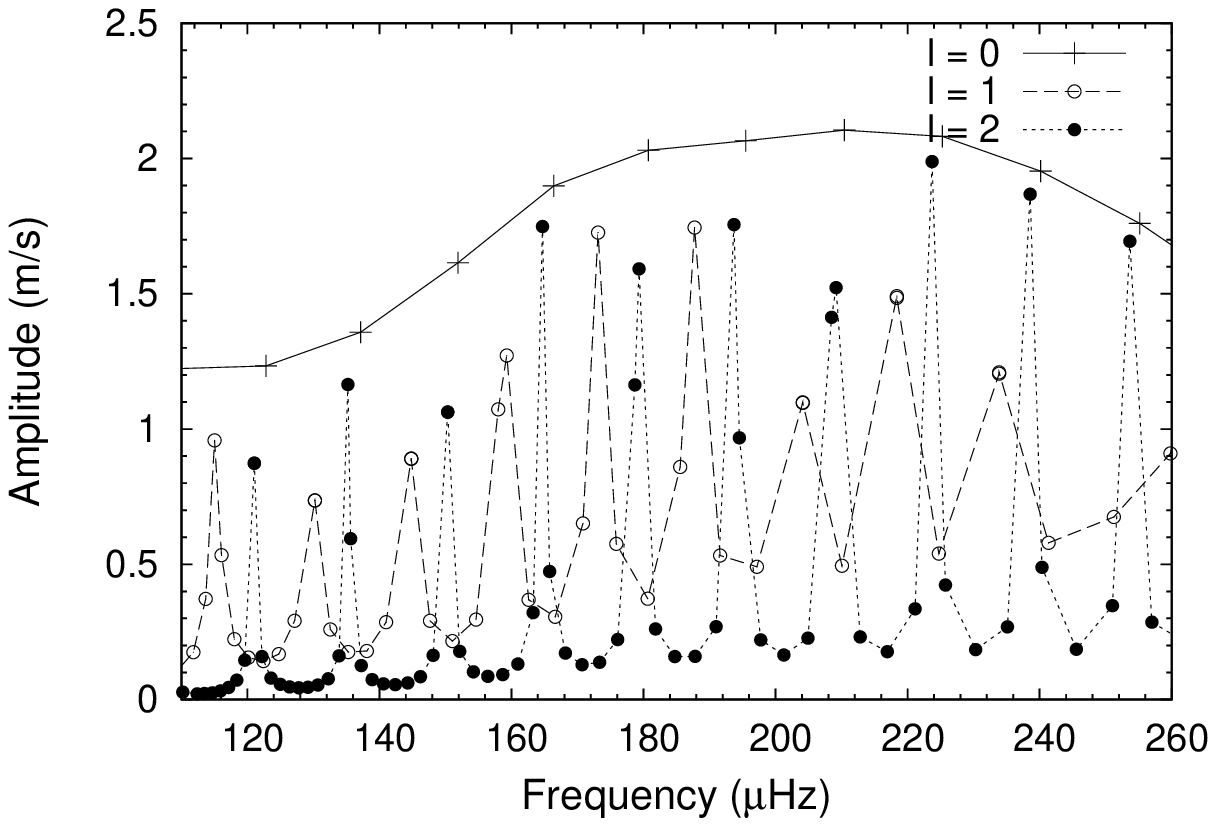}
\includegraphics[width=8.5cm]{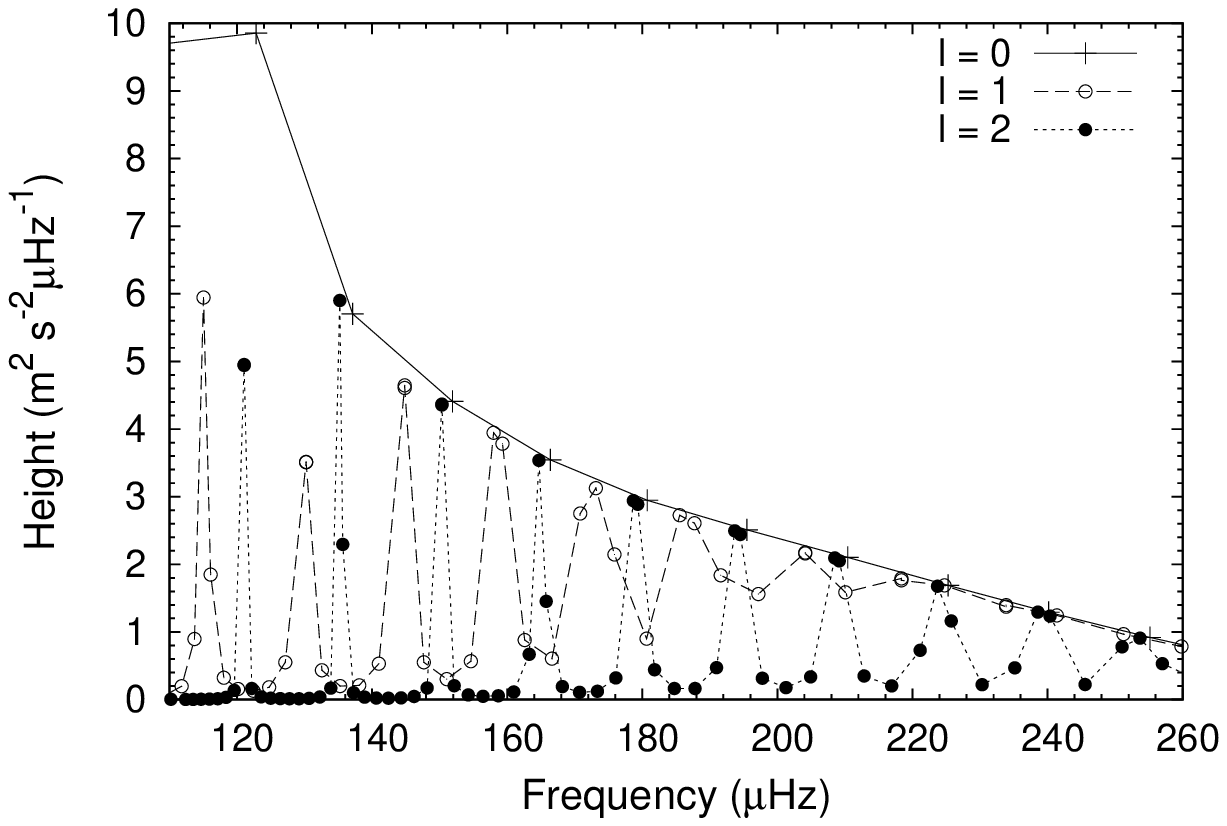}
\caption{Theoretical predictions for $\ell=0$, $1$ and $2$  modes of model~A.
Top panel: theoretical lifetimes; the horizontal line corresponds to our choice
of the resolution limit (75 days).  
Panel~2: lifetimes over inertia ($\tau/I$). Panel~3: amplitudes (m/s).
Panel~4: heights in the PS.}\label{life-A}  
\end{figure} 

\begin{figure}
\includegraphics[width=8.5cm]{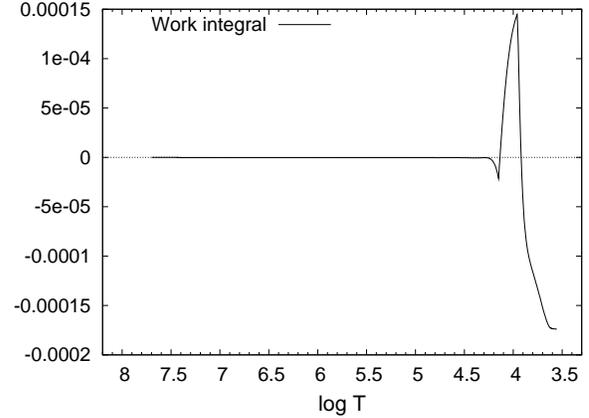}
\caption{Work integral for the mode $\ell=2$, $\nu=201.3\,\mu$Hz trapped in the core of model~A.}
\label{work-A}
\end{figure} 

We first consider the lifetimes of radial modes from the fundamental up to the cut-off frequency.
We also give the resolution limit of 75 days (horizontal line).
As can be seen, we can separate the modes into two sets. From the fundamental to p$_7$
($\nu=123\mu$Hz), the lifetimes are very long and such modes could not be resolved.
On the other side, for larger frequencies the lifetimes of radial modes are below 75 days
and these radial modes have a chance of being resolved. Current observations of red giant
solar-like oscillations show that at least some of the modes can be resolved. We note
also that \cite{Xiong07} predict unstable modes in the lower part of the spectrum.
Depending on the parameter $\beta$ used in our TDC model, they can be stable or unstable.
We restrict our study to the second set of modes, in which radial modes could
be resolved. We propose to call them \emph{solar-like modes} by analogy with the solar case.
Such modes are detected (as shown in the introduction) and their physical interpretation
as stochastic modes has been firmly established.

In the top panel of Fig.~\ref{life-A}, we give the theoretical lifetimes $\tau=1/\eta$ 
obtained for the $\ell=0$, $1$ and $2$  modes of model~A.
According to Eq.~(\ref{workeq}), the mode lifetime $\tau=1/\eta\:\propto\:I$. 
Moreover, we see from Eq.~(\ref{heighteqres}) that the heights in the PS 
are proportional to $(\tau/I)^2$ for resolved modes. This leads us to consider
in the second panel the mode lifetimes divided by the inertia
($\tau/I$). As we showed in Fig.~\ref{in-B1}, the mode trapping leads to an oscillatory
behaviour of the inertia with the large frequency separation as periodicity. 
The oscillatory behaviour of the lifetimes 
is a direct consequence of the oscillations of $I$. But when the ratio between
the two is considered, 
the oscillatory behaviour disappears and the same results are found for radial and non-radial modes. 

In order to interpret this result, we give in Fig.~\ref{work-A} the cumulated
work integral $W(m)=\int_0^m d W/(2\sigma I M)$
for a typical $\ell=2$ mode trapped in the core ($\nu=201.3\,\mu$Hz). The surface value
is $-\eta\,R^{3/2}/(GM)^{1/2}$.  
We see that significant driving and damping only occur in 
the upper part of the convective envelope, but there is no significant radiative
damping in the core. As the radial component of the displacement dominates in the
envelope, the numerator of Eq.~(\ref{workeq}) is the same at a given frequency, 
whatever the value of $\ell$. This explains why $\tau/I$ does not depend on $\ell$.

In the third panel of Fig.~\ref{life-A}, we give the theoretical amplitudes of the modes.
The amplitudes of radial modes are higher because they have less inertia.
The oscillatory behaviour of the inertia due to mode trapping leads to oscillations
of the amplitudes of non-radial modes. 

In the bottom panel of Fig.~\ref{life-A},
we give the theoretical heights in the PS.
We put the resolution limit at $75$~days (horizontal line in top panel).
If the mode lifetime value is larger than this value, Eq.~(\ref{heighteqnores}) is used instead 
of Eq.~(\ref{heighteqres})
to obtain the height. For resolved modes ($H\,\propto\,(\tau/I)^2$), 
the same heights are found for radial and non-radial modes,
except for the visibility factor (integration over the stellar disk), which is not included
in our study. The unresolved non-radial modes with long lifetimes (mainly $\ell=2$ modes
trapped in the core) 
have smaller heights than the closest radial mode.
We emphasize that the predictions for the heights are different
from the amplitudes. Although non-radial modes have smaller amplitudes than radial modes,
the heights in the PS are often of the same order.
Contrary to the amplitudes, the heights are observation dependent. If $T_{\rm obs}$ is smaller, 
more non-radial modes are unresolved and their heights in the PS are smaller. 

Thus, many non-radial modes would be detectable in the PS of stars
like model~A. Moreover, the interaction between the p- and g-mode cavities leads
to many avoided crossings. Hence, the  
resulting frequency spectrum is very complex,
particularly for $\ell=1$ modes and its seismic interpretation is not easy.

\subsection{Model B: intermediate in the red giant branch}
 
\begin{figure}
\includegraphics[width=8.5cm]{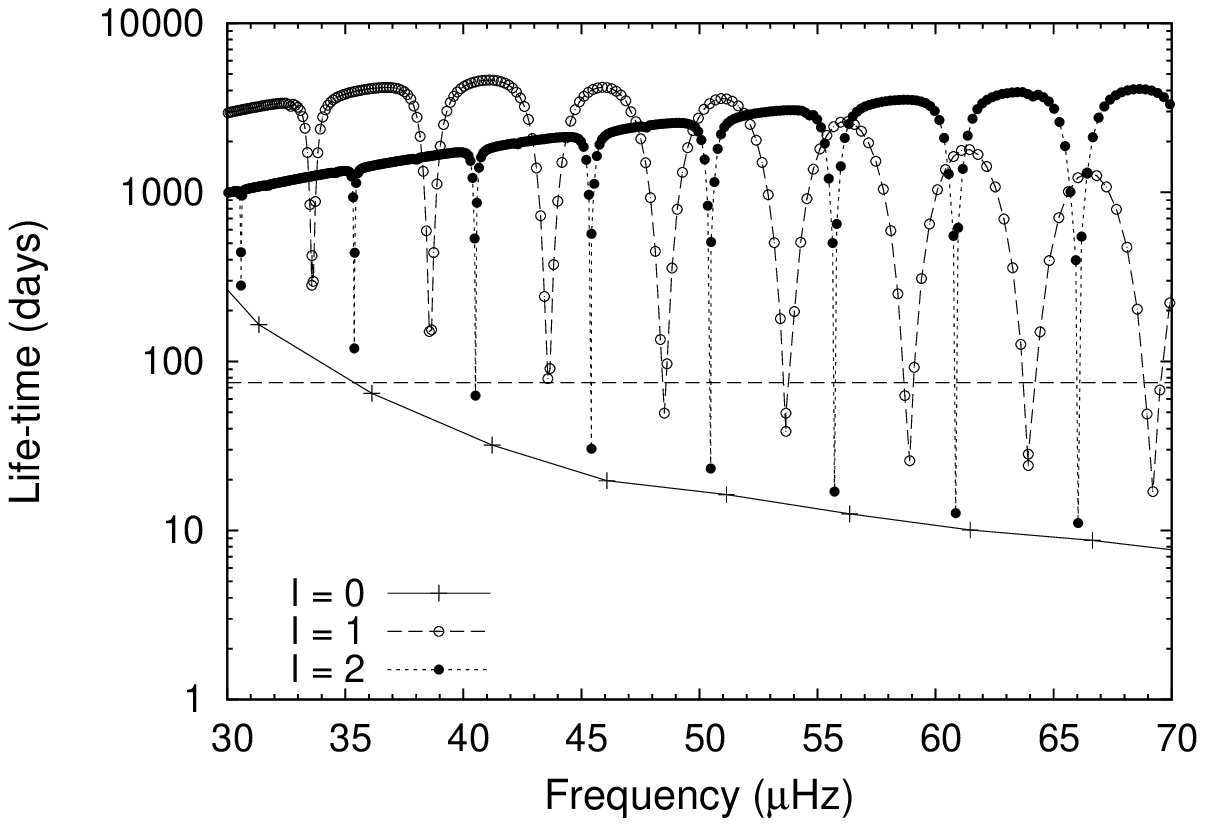}
\includegraphics[width=8.5cm]{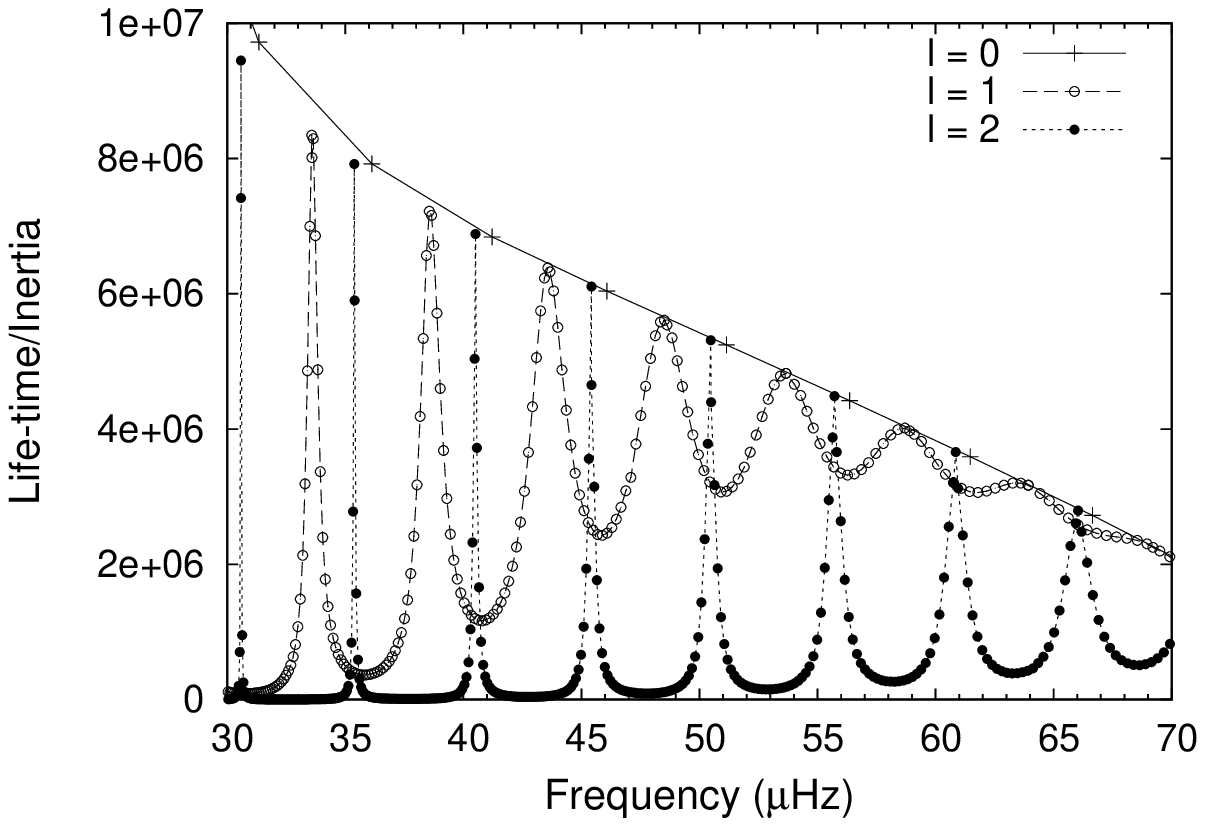}
\includegraphics[width=8.5cm]{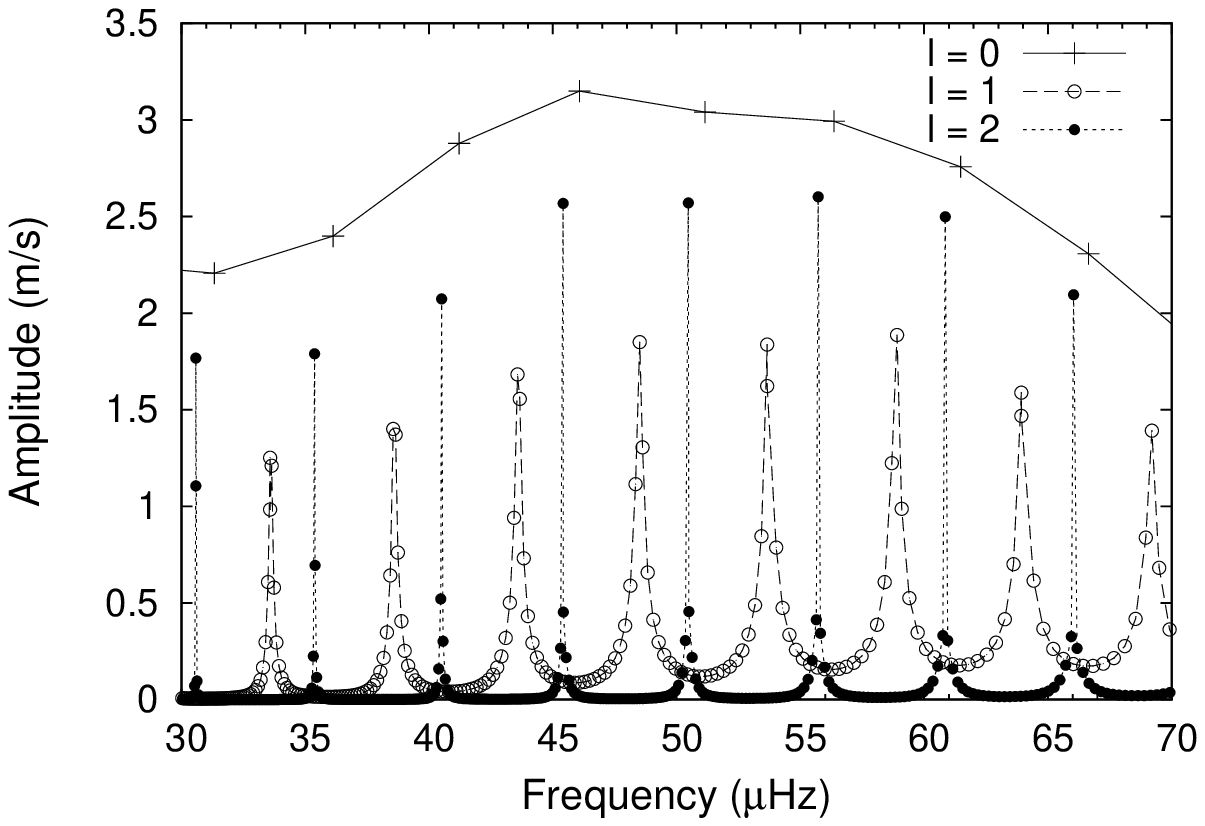}
\includegraphics[width=8.5cm]{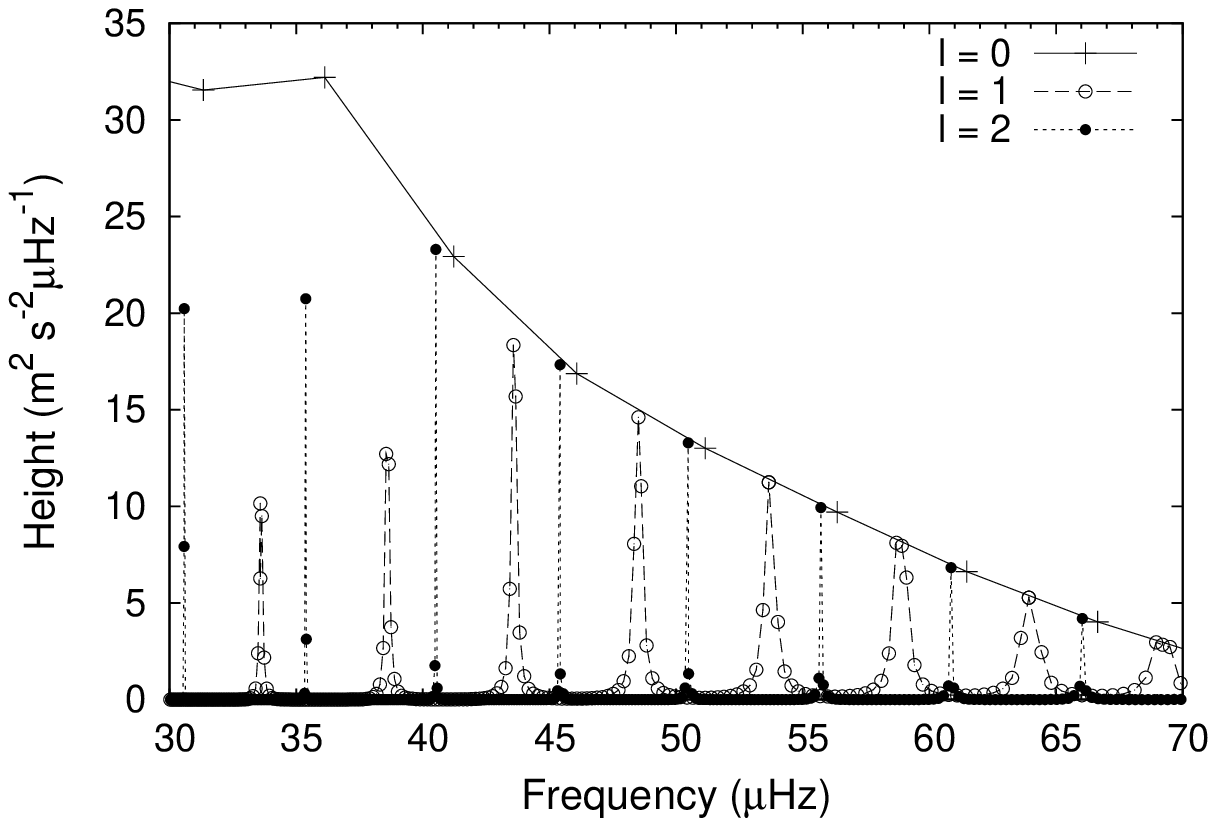}
\caption{Theoretical predictions for $\ell=0-2$  modes of model~B.
Top panel: theoretical lifetimes, 
panel~2: lifetimes over inertia ($\tau/I$), panel~3: amplitudes (m/s),
panel~4: heights in the PS.}
\label{life-B}  
\end{figure} 

\begin{figure}
\includegraphics[width=8.5cm]{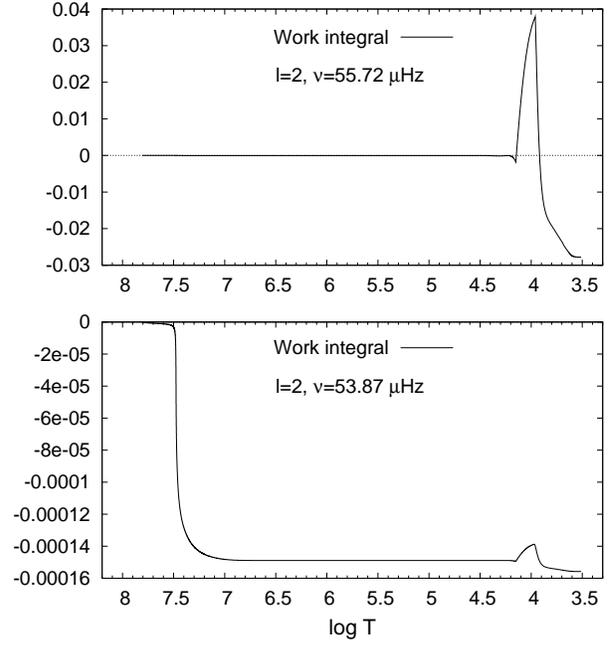}
\caption{Work integrals for the mode $\ell=2$, $\nu=55.72\,\mu$Hz trapped in the envelope
(top panel) and the mode $\ell=2$, $\nu=53.87\,\mu$Hz trapped in the core (bottom panel) 
of model~B.}
\label{work-B}
\end{figure} 

We now present the results obtained for model B that is slightly higher in the red giant
branch (Fig.~\ref{HR}, Table~\ref{models}). In the top panel of Fig.~\ref{life-B},
we give the theoretical lifetimes obtained 
for the $\ell=0$, $1$ and $2$  modes. Again we see the oscillatory behaviour due to
mode trapping. But considering now the ratio between the lifetime and the inertia given in the
2nd panel from the top, we find very different results compared to model~A. Now, oscillatory behaviour 
of this ratio is present for non-radial modes, particularly at low frequencies. 
To understand this, we compare in Fig. ~\ref{work-B}
the work integral for two $\ell=2$ modes, one trapped in the envelope 
($\nu=55.72\,\mu$Hz, top panel) and 
the other trapped in the core ($\nu=53.87\,\mu$Hz, bottom panel). For the mode trapped in the envelope,
the driving and damping occur in the upper part of the convective envelope, as
in model A. But for the mode trapped in the core, significant radiative damping 
occurs around the bottom of the H-burning shell ($\log T\simeq 7.5$). 

In the third panel of Fig.~\ref{life-B}, we give the theoretical amplitudes of radial
and non-radial modes of model~B. The amplitudes of non-radial modes trapped
in the core are very small because of large inertia and radiative damping.
The amplitudes of non-radial modes trapped in the envelope (local maxima in the figure)
are smaller than radial modes but not negligible because of similar inertia and
negligible radiative damping. 

In the bottom panel of  Fig.~\ref{life-B}, we give the 
heights of the modes in the PS. Non-radial modes trapped in the
core have negligible heights and would not be detected. In contrast, non-radial modes trapped in the envelope 
have similar heights compared to radial modes.

Thus, only radial modes and non-radial modes trapped in the envelope
could be observed in the PS of stars like model~B. The groups of non-radial modes
trapped in the envelope are more or less separated from each other by a constant large separation. 
Seismic interpretation of the frequency spectrum of such stars would thus be easier than in model~A.
We notice however that the trapping is not perfect for $\ell=1$ modes. Hence, a small group of
$\ell=1$ modes is detectable around each local maximum. This would have to be taken into account
in any seismic study. Considering observed power spectra, this also means that these groups of modes
would have to be not confused with single large line-widths modes.

\subsection{Model C: high in the red giant branch}

\begin{figure}
\includegraphics[width=8.5cm]{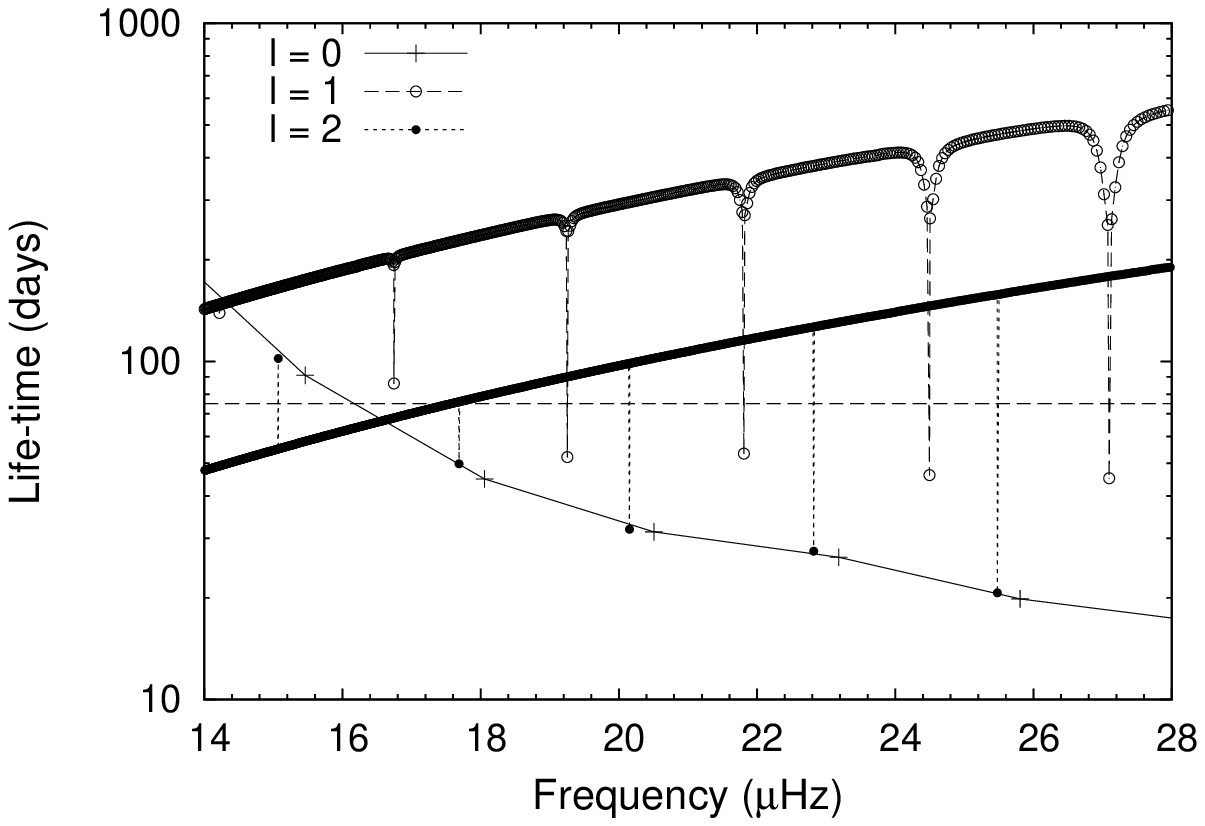}
\includegraphics[width=8.5cm]{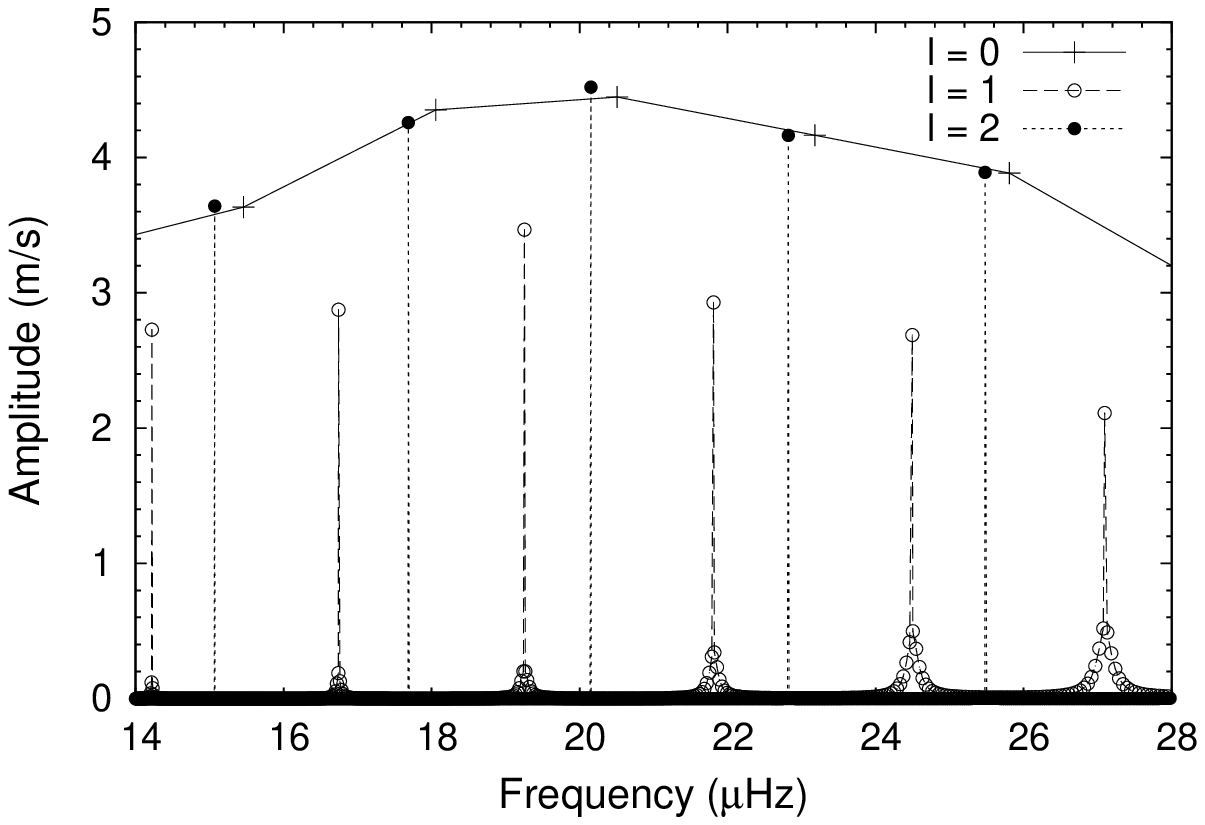}
\includegraphics[width=8.5cm]{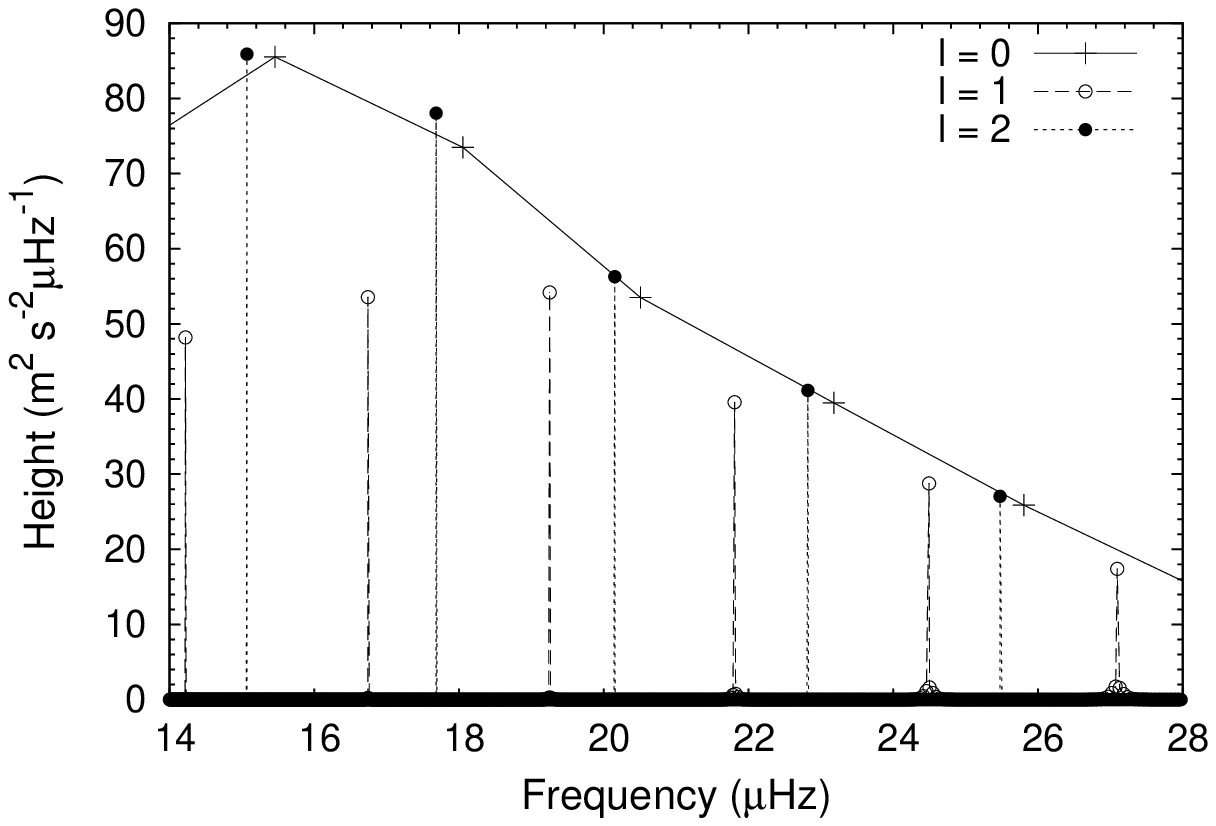}
\caption{Theoretical predictions for $\ell=0-2$  modes of model~C.
Top panel: theoretical lifetimes, 
panel~2: amplitudes (m/s),
panel~3: heights in the PS.}
\label{life-C}  
\end{figure} 

\begin{figure}
\includegraphics[width=8.5cm]{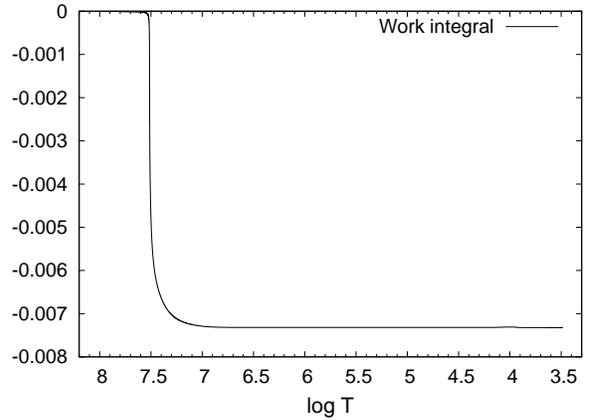}
\caption{Work integral for the mode $\ell=2$, $\nu=22.83\,\mu$Hz of model~C
 (this mode is near a mode trapped in the envelope).}
\label{workC}
\end{figure} 
  
We consider now the results obtained for model C, which is even higher in the red giant
branch than the previous ones (Fig.~\ref{HR}). In the top panel of Fig.~\ref{life-C}, we give the 
theoretical lifetimes. They are very different from those of model~B.
The oscillatory behaviour is no longer present for most $\ell=2$ and low frequency $\ell=1$ modes.
As can be seen in the figure, the transitions to envelope-trapped modes are very sharp; 
these modes, known as Strongly Trapped in the Envelope (STE), are nearly perfectly reflected at the internal 
turning point (near the top of the H-burning shell). 
We give in  Fig.~\ref{workC} the work integral for an $\ell=2$ mode with frequency very close
to but different from an STE mode.
The very high radiative damping illustrated in this figure occurs for all non-radial
modes of this model, except the STE mode. Because of this and the large inertia, 
the g-mode cavity dominates the integrals in the numerator and the denominator of Eq.~(\ref{workeq})
for all non-STE modes.
Hence  Eq.~(\ref{numerator}) and Eq.~(\ref{denominator}) are good approximations for the 
full integrals and the $K$ constant simplifies when the ratio between the two is considered. 
This explains why the lifetimes do not show oscillations in the top panel.
In the second panel of Fig.~\ref{life-C}, we give the theoretical amplitudes. They are 
very small for all non-radial non-STE modes, because of the high
radiative damping and inertia. 
Finally, we give in the bottom panel the theoretical heights in the PS.
All non-radial non-STE modes have negligible heights because of the large radiative damping. 
Thus, only radial modes and non-radial modes completely trapped in the envelope could be 
detected in stars like model~C.

\subsection{Models D and E: H-shell versus He-core burning}
    
\begin{figure}
\includegraphics[width=8.5cm]{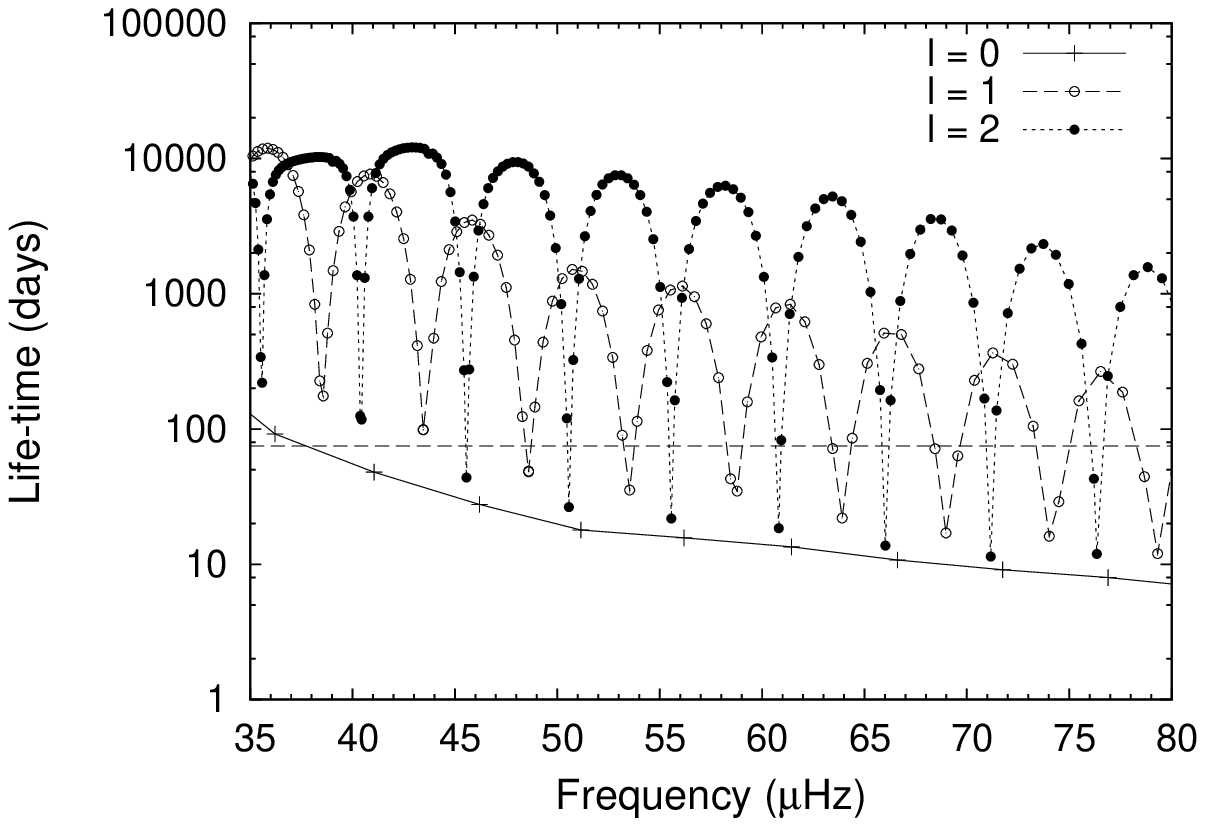}
\includegraphics[width=8.5cm]{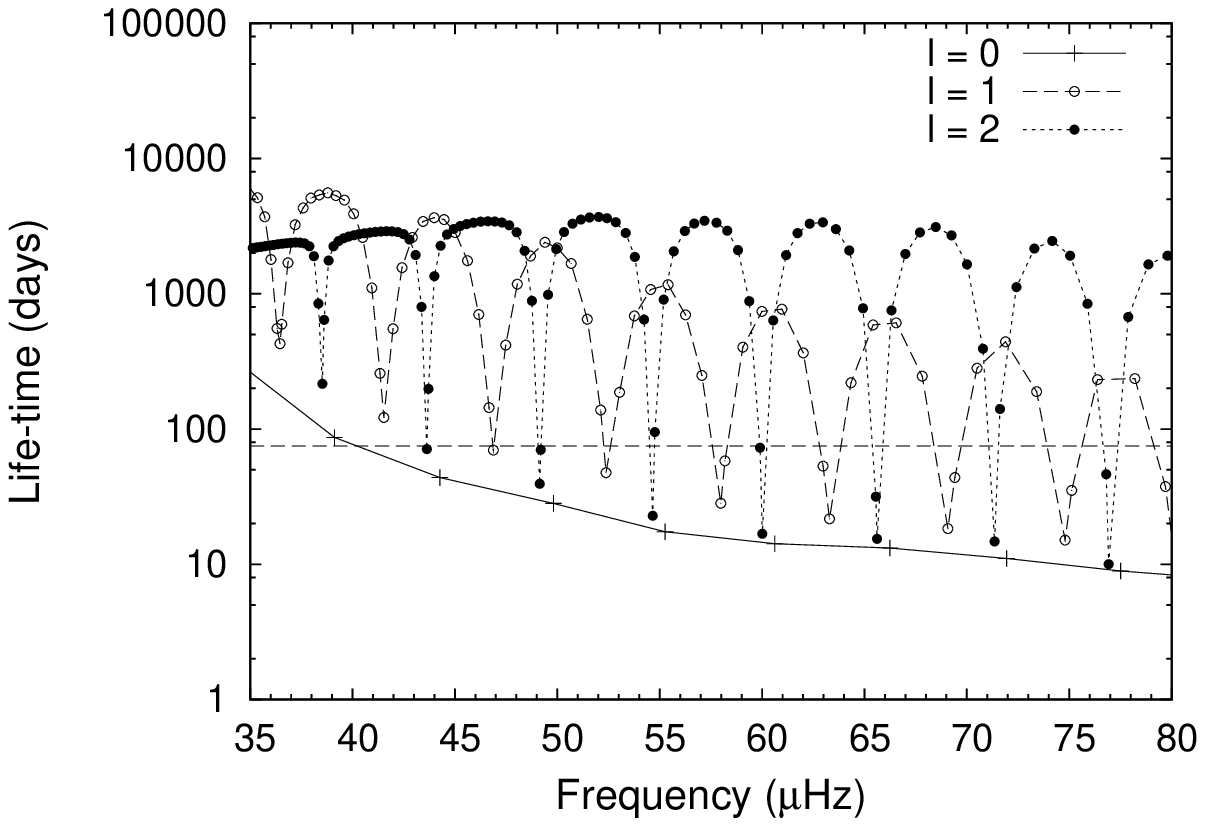}
\caption{Theoretical lifetimes for $\ell=0-2$  modes of model~D (pre-He burning, top panel)
and model~E (He burning, bottom panel).}
\label{life-DE}  
\end{figure} 

\begin{figure}
\includegraphics[width=8.5cm]{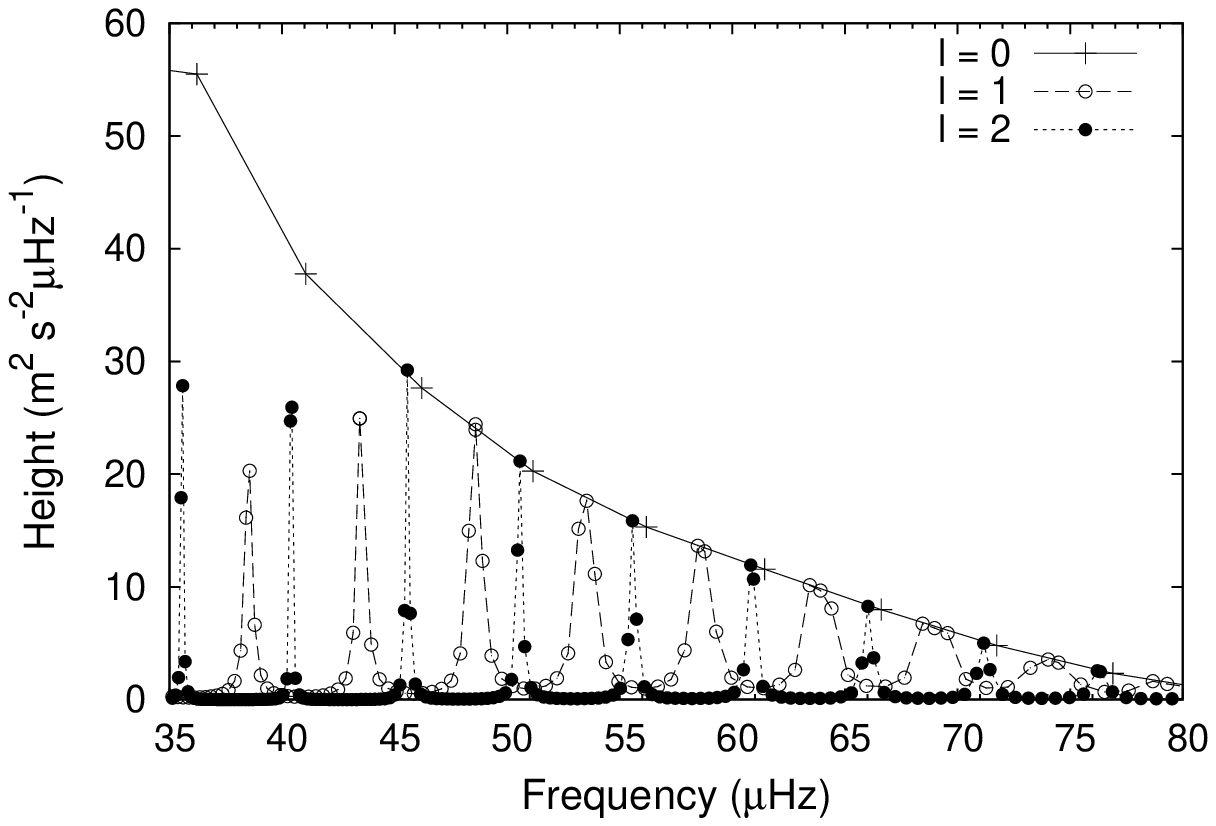}
\includegraphics[width=8.5cm]{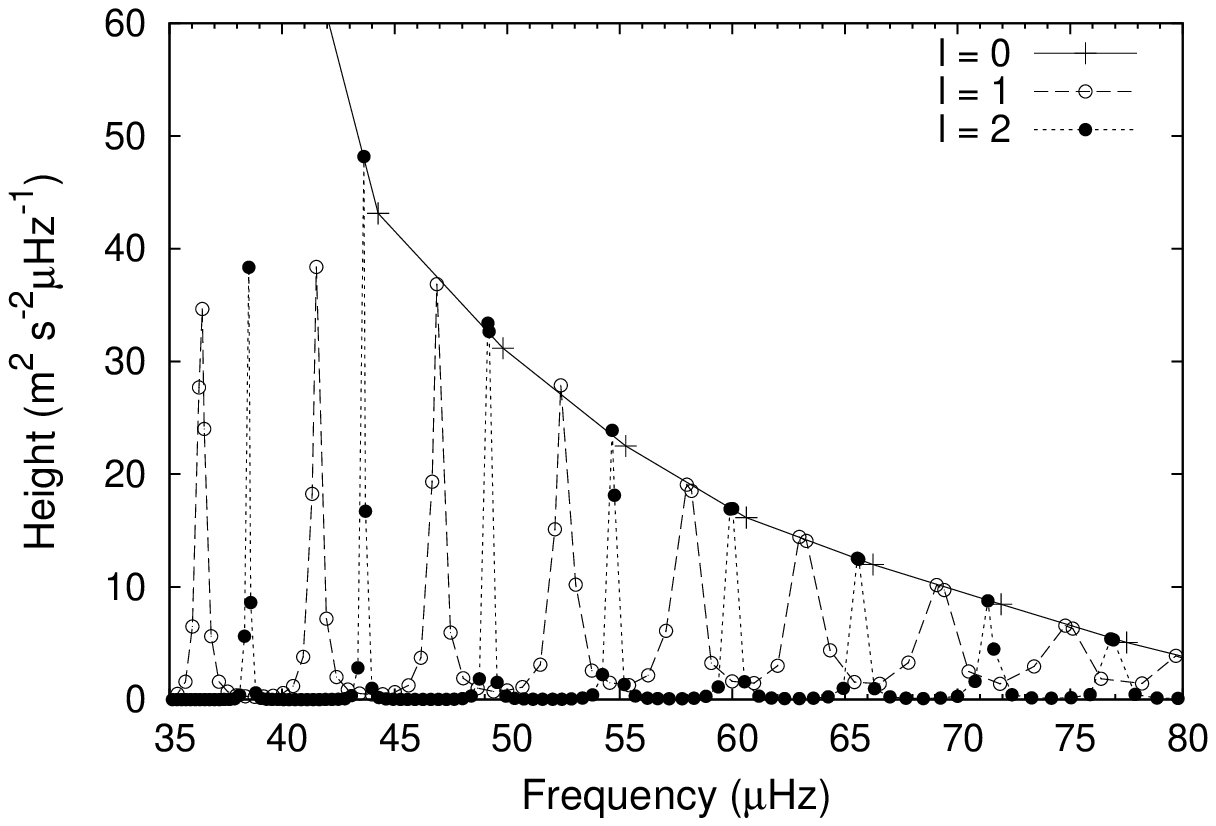}
\caption{Theoretical heights in the PS for $\ell=0,2$ modes
of model~D (top panel) and model~E (bottom panel).}
\label{height-DE}  
\end{figure} 

We now compare the theoretical lifetimes, amplitudes and heights of the two models D and E
with the same luminosity, one  before (model~D) and the other during (model~E) the core helium 
burning phase. 
 
In Fig.~\ref{life-DE}, we give the theoretical 
lifetimes obtained for the $\ell=0-2$  modes of these two models.
The behaviour is similar for the two models. 

In Fig.~\ref{height-DE}, we give the theoretical heights in the PS for these
two models. No significant difference between the two is found except that
the heights in the PS are systematically larger for the He burning model,
because of its higher effective temperature leading to stronger stochastic driving.

\subsection{Uncertainties of the non-adiabatic treatment in the convective envelope}
\label{betachange}

\begin{figure}
\includegraphics[width=8.5cm]{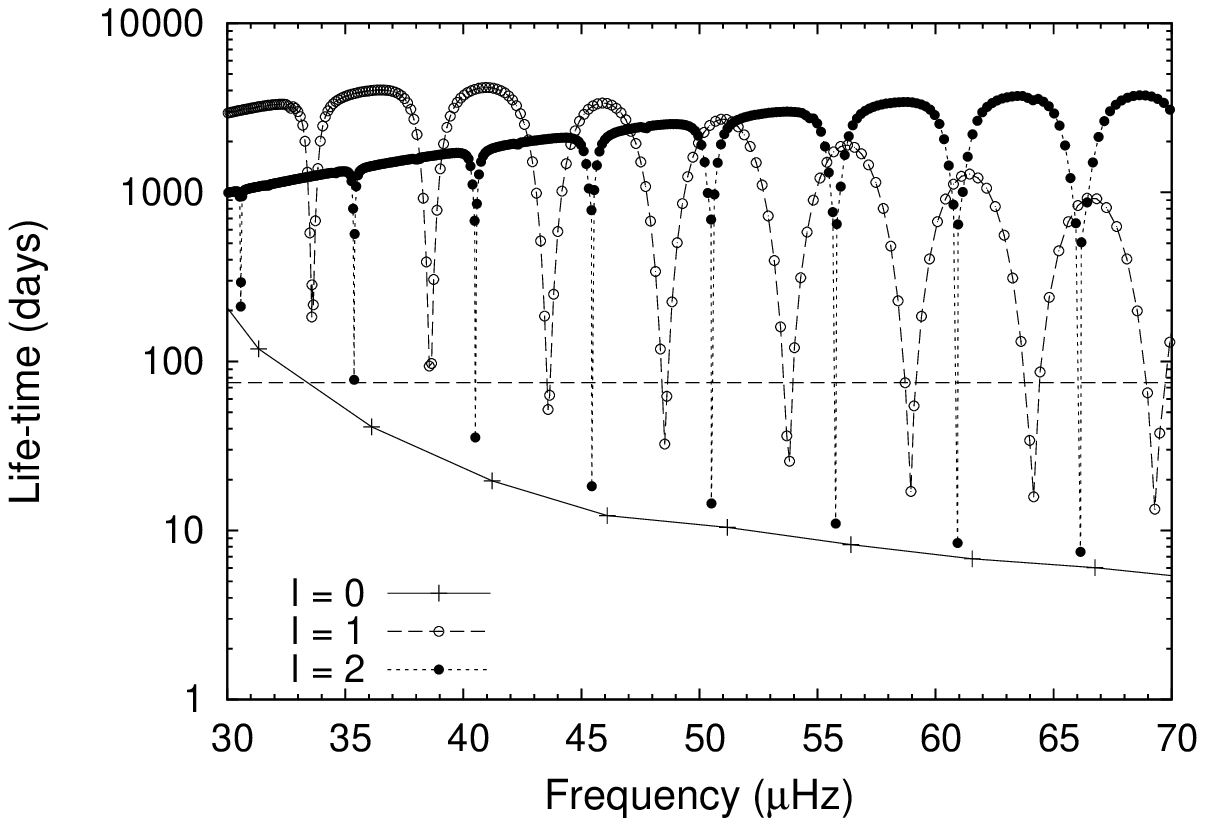}
\includegraphics[width=8.5cm]{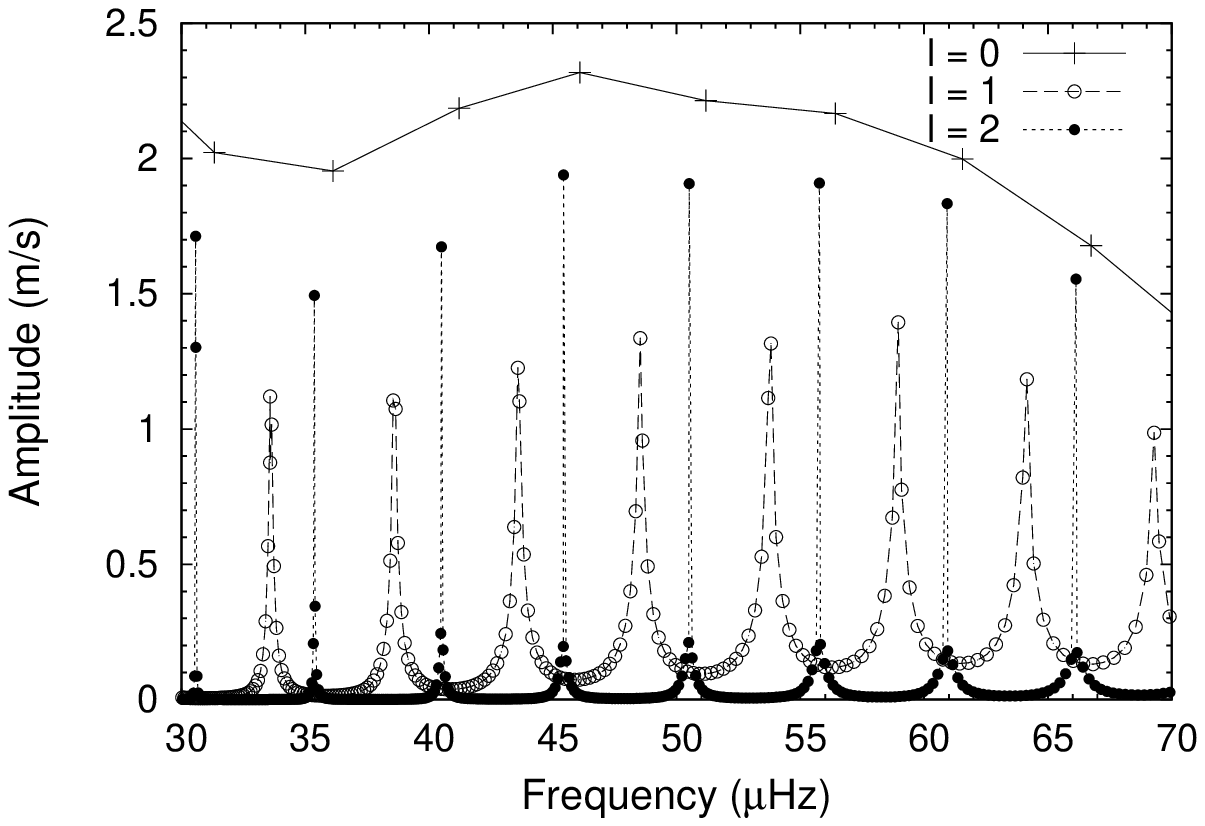}
\caption{Theoretical predictions for $\ell=0$, $1$ and $2$  modes of model~B, 
but with another value of the free parameter of our TDC treatment $\beta=-1.75~i$.
Top panel: theoretical lifetimes, 
bottom panel: heights in the PS.
This figure can be compared with Fig.~\ref{life-B} for which $\beta=-2~i$.}\label{life-Bbet}  
\end{figure} 

As discussed in Sects.~\ref{TDCsec} and \ref{nonadsec}, the time-dependent interaction between convection 
and oscillations is subject to large uncertainties. In the TDC treatment of
G05, a free complex parameter $\beta$ was introduced
in the perturbed closure equations. In the previous computations, we adopted $\beta=-2\,i$. 
We now redo the computations with $\beta=-1.75\,i$. Larger changes of $\beta$ would
give unstable modes in the considered range of frequencies,
which is not realistic in view of the typical observed amplitudes of red giant high order
p-modes.

With $\beta=-1.75~i$, the convective damping in the upper part of the convective envelope
is larger. As shown in Fig.~\ref{life-Bbet}, this results in smaller lifetimes and amplitudes
of the modes trapped in the envelope compared to Fig.~\ref{life-B}. 
In the upper part of the envelope, the eigenfunctions do not depend significantly
on the degree $\ell$ of the modes. Hence the uncertainties associated with the TDC treatment
affect radial and non-radial modes trapped in the envelope in the same way.

\subsection{Non-linear effects?}

Our study is linear, raising the question of whether such an approximation is valid here.
Considering first the amplitudes of the displacement at the photosphere,
we find typical values about $200$ times smaller than the pressure scale height
for detectable modes.
Hence non-linear effects in the superficial layers are expected to be small.
We consider now the oscillations in the g-mode cavity.
We can confidently neglect non-linear terms if 
$|{\rm v}_r\partial \vec{{\rm v}}/\partial r|<<|\partial\vec{{\rm v}}/\partial t|$.
In a cavity with short wavelength oscillations, this condition is equivalent to 
$|\xi_r|<<1/k_r$.  In the g-mode cavity of envelope trapped modes, $|\xi_r|$ is typically 
about $10^4$ times smaller than $1/k_r$.
Hence the risk of wavebreaking due to non-linear effects
is negligible in the core.

\section{Conclusions}

We have determined the theoretical lifetimes, amplitudes and heights in the PS  
of radial and non-radial modes of several red giant models. 
The predictions appear to be very different depending on the evolutionary status of the star. 
What mainly matters is the density contrast between the He core and the envelope. 
During evolution, it increases considerably as the He core contracts and the envelope expands. 
Moreover the Kelvin-Helmholtz time decreases because of the increasing luminosity.
The consequence of these two effects is an increasing importance of the radiative damping
of non-radial modes as the luminosity of the star increases.
We can encounter very different cases depending on the evolutionary status of the star:

\begin{itemize}
\item Case A: Subgiants and/or low luminosity red giants.\\
Our model representative of the class is model~A at the bottom of the giant branch 
for $2 M_{\sun}$. In these stars, the density contrast is not large enough to imply
significant radiative damping of non-radial solar-type p-modes
($R_c/R= 0.0075$ in model A). The amplitudes of non-radial modes are smaller
than radial modes because of larger inertia. However, if we consider instead the heights
in the PS, they are similar for radial and non-radial modes as long
as they can be resolved. As the heights are used as detection criteria
in a Fourier analysis, many non-radial modes could be detected in the PS of such stars. 
The interaction between the p- and g-mode cavities leads to many avoided crossings. 
The resulting frequency spectrum is thus very complex. Asteroseismology of such stars would be challenging.

\item Case B: Intermediate models in the red giant branch.\\
Our models representative of this class are model~B, D and E with greater luminosity
compared to the previous case  ($\log(L/L_{\sun})=1.8-2$). In these stars
the radius ratio between the core and the envelope is smaller ($R_c/R= 0.003$ in model B).
As a consequence, the radiative
damping around the bottom of the H-burning shell becomes large enough to destroy the 
non-radial modes trapped in the core. Only the non-radial modes trapped in the envelope
and the radial modes can survive and have similar heights in the PS. 
The non-radial modes
trapped in the envelope are more or less separated from each other by a constant large separation,
allowing us to build echelle diagrams. 
However, the trapping is not perfect for $\ell=1$ modes. Hence, a small group of
$\ell=1$ modes is detectable around each local maximum in the PS corresponding to an envelope trapped mode.
Such structures would not have to be interpreted as single large line-width modes. 
Seismic interpretation of the frequency spectrum of
such stars would be easier than in the previous case.

\item Case C: High luminosity red giants.\\
Our model representative of this class is model~C with $M=2 M_{\sun}$, $\log(L/L_{\sun})=2.1$.
This case is not very different from the previous one, except that
the density contrast between the core and the envelope is very large ($R_c/R= 0.002$ in
model~C). As a consequence, the radiative damping is stronger in the core and destroys
all non-radial modes except those strongly trapped in the envelope. These modes and the radial modes 
are the only ones that could be detected in these stars. 
The trapping is efficient enough in $\ell=1$ modes and only \emph{single} envelope trapped modes
are detectable. Seismic interpretation of the frequency spectrum of
such stars would also be easier.

\end{itemize}

We also notice that the duration of observation is very important, as it determines
the frontier between resolved and unresolved modes.
Non-radial modes have always much longer lifetimes
($\tau$) than radial modes because of their greater inertia. 
If $\tau \gtrsim T_{\rm obs}/2$, the modes cannot be resolved
and have much smaller heights in the PS. 
With too short observations, 
as is typically the case with ground based observations,
only radial and strongly trapped non-radial modes have a chance of being detected.
On the other hand, future space missions will provide longer observations: e.g. 5 years for Kepler,
3 years for Plato, increasing the ability to detect more non-radial modes.

Asteroseismology of red giants is a very promising field as the theoretical predictions
depend strongly on the physical characteristics of both their deep and superficial
layers.

\begin{acknowledgements} 
The works of K. Belkacem were financially supported throught 
a postdoctoral fellowship from the 
``Subside f\'ed\'eral pour la recherche 2009'', University of Li\`ege.
We thank Prof. W. Dziembowski for his useful advices and remarks.
\end{acknowledgements}

\bibliographystyle{aa}
\bibliography{1713.bib}

\end{document}